\documentclass[a4paper,11pt]{article}

\usepackage[a4paper,margin=25mm]{geometry}
\usepackage{amsmath,amssymb,amsfonts,mathtools,bm,array}
\usepackage{xcolor}
\definecolor{linkblue}{RGB}{0,50,232}
\definecolor{equationpink}{RGB}{232,0,120}
\usepackage[colorlinks=true,linkcolor=linkblue,citecolor=linkblue,urlcolor=linkblue]{hyperref}
\makeatletter
\renewcommand{\eqref}[1]{%
 \textup{\hyperref[#1]{\textcolor{equationpink}{\tagform@{\ref*{#1}}}}}%
}
\makeatother
\hypersetup{
 pdftitle={Spectral Fractional Bosonic Strings: Exact Polyakov Measure and the Critical Dimension},
 pdfauthor={Tianhao Wu and Gabriele La Nave}
}
\usepackage{microtype}
\allowdisplaybreaks
\newenvironment{widetext}{}{}
\newcommand{\dd}{\mathrm d}
\newcommand{\e}{\mathrm e}
\newcommand{\ii}{\mathrm i}
\newcommand{\Tr}{\mathrm{Tr}}
\newcommand{\tr}{\mathrm{tr}}
\newcommand{\mc}{\mathcal}
\newcommand{\vev}[1]{\left\langle #1 \right\rangle}
\newcommand{\Ran}{\mathrm{Ran}}
\newcommand{\Area}{\mathrm{Area}}
\newcommand{\Vol}{\mathrm{Vol}}
\newcommand{\gh}{\mathrm{gh}}

\newcommand{\doilink}[1]{\href{https://doi.org/#1}{\textcolor{linkblue}{doi:#1}}}
\newcommand{\arxivlink}[1]{\href{https://arxiv.org/abs/#1}{\textcolor{linkblue}{arXiv:#1}}}

\numberwithin{equation}{section}

\begin{document}

\title{Spectral Fractional Bosonic Strings: Exact Polyakov Measure and the Critical Dimension}

\author{
Tianhao Wu$^{1}$\thanks{\href{mailto:twu49@illinois.edu}{twu49@illinois.edu}}
\quad and \quad
Gabriele La Nave$^{1}$\\[0.6em]
\small $^1$Department of Physics, University of Illinois Urbana-Champaign, Urbana, Illinois 61801, USA}
\date{}

\maketitle

\begin{abstract}
Fractional Laplacians encode anomalous scaling and covariant nonlocal response in quantum matter and holographic boundary systems. We place the corresponding scalar spectral theory on a dynamical closed worldsheet. A metric-dependent intertwiner maps the fractional quadratic form to that of $D$ free bosons, while its zeta Jacobian changes the matter determinant weight from $D$ to $sD$. The prime-determinant transformation includes the zero-mode area factor and determines both the moduli density and the Weyl coefficient $c_{\rm det}=sD$. Weyl cancellation therefore selects
$D_\ast=26/s$. On the integer critical locus $1\leq D\leq26$, $N=26-D$ spectator bosons turn the residual determinant into a local critical CFT. Its fixed-area genus-$g$ vacuum measure is that of 26 free bosons, its torus trace is modular invariant, its plumbing channels are state resolved, and its BRST charge is nilpotent. The $D$ distinguished coordinates remain spectral composites; their pullback yields the Koba--Nielsen amplitudes and massless vertex conditions. Target backgrounds coupled to this distinguished sector retain the ordinary one-loop tensor beta functions, with dilaton deficit $sD-26$ and the corresponding string-frame action. Worldsheet gravity thus turns the fractional exponent into a quantum-geometric consistency parameter.
\end{abstract}

\section{Introduction}

Fractional Laplacians describe anomalous scaling without abandoning geometric covariance. In quantum matter they govern long-range response and transport~\cite{Pippard1953,LaNaveRMP2019,GubserNonlocalSigma2019,HeydemanNonlocalQED2020}; in holography they arise as Dirichlet-to-Neumann operators for boundary currents~\cite{LaNavePhillips2019}. String theory adds a decisive element: the metric defining the operator is itself integrated. The fractional sector must then enter the Weyl and moduli measures of the worldsheet.

Quantum-critical transport gives a concrete motivation. Fractional electromagnetism assigns noncanonical dimensions to charge and current while preserving generalized gauge structure, organizing both Pippard response and anomalous holographic boundary currents~\cite{Pippard1953,LaNaveRMP2019,LaNavePhillips2019,HartnollKarch2015}. Hyperscaling-violating geometries, nonlocal sigma models, and nonlocal QED place the same mechanism in controlled quantum regimes~\cite{GouterauxKiritsis2011,Karch2014,GubserNonlocalSigma2019,HeydemanNonlocalQED2020}. Their common ingredient is a covariant function of a geometric Laplace-type operator.

On a worldsheet, the spectral exponent controls both matter scaling and the metric dependence of the functional measure. Spectral functional calculus is global and diffeomorphism covariant on a compact surface, so it poses a precise question: what Weyl weight does a fractional embedding field acquire once the worldsheet metric fluctuates? Earlier fractional-string models developed flat-worldsheet dynamics, fractional symmetries, and asymptotic states~\cite{DiazGiusti2018,Diaz2020}; fixed-background Gaussian theories are related to local ones by a nonlocal Hilbert-space transform~\cite{BasaClassification}. Promoting the metric to a Polyakov variable makes this transform move over metric space, and its Jacobian becomes dynamical data.

Let $h$ be a metric on a smooth closed Riemann surface and let $X=(X^1,\ldots,X^D)$ denote the spectral embedding fields. Removing the constant mode from the positive scalar Laplace--Beltrami operator $-\Delta_h$ gives $A_h>0$. Its real power $A_h^s$, defined by self-adjoint functional calculus, generates the compact theory for every $s>0$. In the flat regime $0<s<1$, the same operator is the boundary response of a local weighted extension.

The key object is the metric-dependent map $\mc I_h$. It fixes the target translation mode and sends the fractional form to the ordinary free-boson form. Defining the $X$ measure in $L^2(h)$ eigenmode coordinates and writing $Y=\mc I_hX$ gives
\begin{equation*}
\begin{aligned}
Y&=\mc I_hX,\\
\mc D_hX&=
 \left[\det{}'(\ell^2A_h)\right]^{-(s-1)D/2}\mc D_hY .
\end{aligned}
\end{equation*}
Here $\ell$ fixes the determinant scale and the prime omits the scalar zero mode. The nonzero modes generate the determinant; the constant mode passes unchanged through the map.

The metric dependence of $\mc I_h$ is the source of the effect. Its Jacobian changes the determinant exponent from $D$ to $sD$; the Polyakov--Alvarez law then turns $s$ into a determinant-line Weyl weight and a density over moduli space. Cancellation against the $bc$ ghosts selects
\begin{equation*}
c_{\rm det}=sD,\qquad
sD=26\quad\Longrightarrow\quad D_\ast=\frac{26}{s}.
\end{equation*}
The local completion below realizes this geometric weight as a $c_m=26$ operator theory.

On the integer critical locus $1\leq D\leq26$, $N=26-D$ spectator bosons turn the residual determinant into a local tensor-product CFT. Its fixed-area genus-$g$ vacuum measure equals the 26-boson measure, its genus-one trace is modular invariant, and its plumbing channels run over $\mc H_Y\otimes\mc H_\chi$. The resulting BRST charge is nilpotent. The $D$ distinguished coordinates remain spectral composites of $Y$, preserving the target zero mode and reproducing the Koba--Nielsen amplitudes and massless closed-string vertices.

Background fields $G_{\mu\nu}$, $B_{\mu\nu}$, and $\Phi$ couple to the distinguished $Y$ coordinates, with the spectator CFT left free. The one-loop tensor beta functions keep their ordinary form, while the dilaton deficit becomes $sD-26$ and follows from the corresponding string-frame action.

Sections~\ref{sec:spectral}--\ref{sec:gaugefix} develop the spectral matter measure and follow its Jacobian through Weyl transformation and conformal gauge fixing. Sections~\ref{sec:completion}--\ref{sec:observables} build the critical operator theory and its observables; the appendices collect the determinant, localization, BRST, sewing, and background-field calculations.

\section{Spectral fractional matter on compact worldsheets}\label{sec:spectral}

\subsection{Positive scalar operator and form domain}

Let $(\Sigma,h)$ be a smooth compact connected oriented Riemann surface without boundary, with volume form $\dd\mu_h=\dd^2\sigma\sqrt h$. The scalar Laplace--Beltrami operator acts on test functions as
\begin{equation}
\Delta_h f=\frac{1}{\sqrt h}\partial_a\!\left(\sqrt h\,h^{ab}\partial_b f\right),\qquad -\Delta_h\ge0 .
\end{equation}
The operator $-\Delta_h$ is essentially self-adjoint on $C^\infty(\Sigma)$. Choosing a real orthonormal eigenbasis, its spectrum is
\begin{eqnarray}
&&-\Delta_h\varphi_n =\lambda_n\varphi_n  \;\;, \;\;  \langle\varphi_m, \varphi_n\rangle_{L^2(h)} =\delta_{mn}, \nonumber \\
&& \;\; 0=\lambda_0<\lambda_1\le\lambda_2\le\cdots .
\end{eqnarray}
The zero eigenspace is spanned by the normalized constant
\begin{equation}
\varphi_0=\Area[h]^{-1/2},
\qquad
\Area[h]=\int_\Sigma \dd\mu_h.
\end{equation}
We separate this mode by the orthogonal projectors
\begin{align}
P_0(h)=|\varphi_0\rangle\langle\varphi_0|, \qquad
P_\perp(h)=1-P_0(h), \qquad
\Ran(P_\perp)&=\varphi_0^\perp\subset L^2(\Sigma).
\end{align}
Equivalently, $f\in\Ran(P_\perp)$ precisely when $\int_\Sigma\dd\mu_h f=0$. The intrinsic nonzero-mode field is an element of the quotient
\begin{equation}
\mathfrak H_h^r=H^r(\Sigma,h)/\mathbb R ,
\label{eq:quotient-space-long}
\end{equation}
and $P_\perp(h)$ selects its unique $h$-mean-zero representative. Pullback by a diffeomorphism preserves the class modulo constants, while the target translation coordinate forms the complementary finite-dimensional mode.
The positive operator used throughout is
\begin{equation}
A_h=P_\perp(-\Delta_h)P_\perp\quad\text{on }\Ran(P_\perp).
\end{equation}
On this subspace $A_h$ is strictly positive, with $\langle f,A_hf\rangle\ge\lambda_1\|f\|^2$, and hence induces an invertible positive operator on $\mathfrak H_h^0$. The projector and the $L^2(h)$ metric both vary with $h$; together they determine the metric response of the spectral measure.

For $s>0$ we define $A_h^s$ by self-adjoint functional calculus,
\begin{equation}
A_h^s f=\sum_{n\ge1}\lambda_n^s f_n\varphi_n,
\qquad
f_n=\langle\varphi_n,f\rangle_{L^2(h)} .
\end{equation}
As an unbounded operator, $A_h^s$ has domain
\begin{equation}
\mc D(A_h^s)=\left\{f\in\Ran(P_\perp):\sum_{n\ge1}\lambda_n^{2s}|f_n|^2<\infty\right\}.
\end{equation}
The quadratic form in the action uses $A_h^{s/2}$, hence its natural form domain is
\begin{equation}
\mc D(A_h^{s/2})=\left\{f\in\Ran(P_\perp):\sum_{n\ge1}\lambda_n^s |f_n|^2<\infty\right\}.
\label{eq:formdomain-long}
\end{equation}
By elliptic regularity on a compact smooth surface this is the zero-mode-free Sobolev space $H^s(\Sigma)\cap\Ran(P_\perp)$, with equivalent norms~\cite{Seeley1967,ReedSimon}. Spectral cutoff followed by zeta continuation defines the Gaussian integral. Its Cameron--Martin space is Eq.~\eqref{eq:formdomain-long}; typical fields have the lower Sobolev regularity set by the covariance.

\subsection{Fractional action and fixed-metric Gaussian theory}

For the real embedding fields $X:\Sigma\to\mathbb R^D$, write
$\widetilde X^\mu=P_\perp(h)X^\mu$ and choose
$g_{\mu\nu}^{\rm E}=\operatorname{diag}(+1,\ldots,+1)$. The reference
length $\ell>0$ makes the closed quadratic form dimensionless:
\begin{align}
S_s[X;h]
&:=\frac{\ell^{2s-2}}{4\pi\alpha'}\sum_{\mu=1}^D
 \|A_h^{s/2}\widetilde X^\mu\|^2_{L^2(h)} =\frac{\ell^{2s-2}}{4\pi\alpha'}\,
 g_{\mu\nu}^{\rm E}
 \left\langle A_h^{s/2}\widetilde X^\mu,
 A_h^{s/2}\widetilde X^\nu\right\rangle_{L^2(h)}
 \nonumber\\
&=\frac{\ell^{2s-2}}{4\pi\alpha'}\,
 \int_\Sigma\dd^2\sigma\,\sqrt{h(\sigma)}
 g_{\mu\nu}^{\rm E} 
 \bigl(A_h^{s/2}\widetilde X^\mu\bigr)(\sigma)
 \bigl(A_h^{s/2}\widetilde X^\nu\bigr)(\sigma)
 \nonumber\\
&=\frac{\ell^{2s-2}}{4\pi\alpha'}\,
 g_{\mu\nu}^{\rm E}\sum_{n=1}^{\infty}
 \lambda_n^sX_n^\mu X_n^\nu .
\label{eq:fractional-action-long}
\end{align}
The mode coefficients are
\begin{align*}
X_n^\mu&:=\langle\varphi_n,X^\mu\rangle_{L^2(h)} =\int_\Sigma\dd^2\sigma\,\sqrt{h(\sigma)}\,
\varphi_n(\sigma)X^\mu(\sigma),\qquad n\geq1 .
\end{align*}
These are the nonzero-mode coefficients of $\widetilde X^\mu$, so
$(A_h^{s/2}\widetilde X^\mu)(\sigma)
=\sum_{n\geq1}\lambda_n^{s/2}X_n^\mu\varphi_n(\sigma)$ in $L^2(h)$.
Parseval's identity gives the final line of
Eq.~\eqref{eq:fractional-action-long}. It weighs the $n$th mode by
$\lambda_n^s$, and its finite-action space is Eq.~\eqref{eq:formdomain-long}.

At $s=1$ the reference length drops out. On the closed worldsheet, $A_h=P_\perp(-\Delta_h)P_\perp$ and
$\partial_a\widetilde X^\mu=\partial_aX^\mu$ give
\begin{align*}
S_1[X;h]
&=\frac{1}{4\pi\alpha'}\sum_{\mu=1}^D\sum_{n=1}^{\infty}
\lambda_n\bigl(X_n^\mu\bigr)^2 =\frac{1}{4\pi\alpha'}\int_\Sigma\dd^2\sigma\,\sqrt h\,
h^{ab}g_{\mu\nu}^{\rm E}\partial_aX^\mu\partial_bX^\nu ,
\end{align*}
which is the Euclidean Polyakov matter action
$S_{\rm P}^{(E)}[X;h]$.

The full field decomposes as
\begin{equation}
 X^\mu=x_0^\mu+\widetilde X^\mu,
 \qquad P_0X^\mu=x_0^\mu,
 \qquad \widetilde X^\mu=P_\perp X^\mu .
\label{eq:Xsplit-long}
\end{equation}
Thus $A_h^s$ acts on $\widetilde X^\mu$, while $x_0^\mu$ remains the target translation coordinate.

At a finite spectral cutoff $M$, the nonzero-mode action is
\begin{equation}
S_{s,\perp}^{(M)}
=\frac{1}{4\pi\alpha'}\sum_{\mu=1}^D\sum_{n=1}^M
\ell^{2s-2}\lambda_n^s\bigl(X_n^\mu\bigr)^2 .
\label{eq:finite-mode-action-long}
\end{equation}
At the same cutoff, define the $L^2(h)$-ultralocal measure on the
nonzero-mode sector by
\begin{equation*}
\mc D_{h,\perp}^{(M)}X
:=\prod_{\mu=1}^D\prod_{n=1}^M\dd X_n^\mu .
\end{equation*}
\begin{align}
Z_{m,\perp}^{(s,M)}[h]
&:=\int\mc D_{h,\perp}^{(M)}X\,
\e^{-S_{s,\perp}^{(M)}}  =\prod_{\mu=1}^D\prod_{n=1}^M
\int_{\mathbb R}\dd X_n^\mu\,
\exp\!\left[
-\frac{\ell^{2s-2}\lambda_n^s}{4\pi\alpha'}
\bigl(X_n^\mu\bigr)^2
\right]\nonumber\\
&=C'_M\prod_{n=1}^M
\left(\ell^{2s-2}\lambda_n^s\right)^{-D/2}\nonumber\\
&=C_M\prod_{n=1}^M
\left(\ell^{2s}\lambda_n^s\right)^{-D/2},
\label{eq:finite-gaussian-product-long}
\end{align}
where $C'_M=(4\pi^2\alpha')^{DM/2}$ and
$C_M=\ell^{DM}C'_M=(4\pi^2\alpha'\ell^2)^{DM/2}$.
The $L^2(h)$ mode measure therefore retains the geometry through $\lambda_n[h]$. Its zeta limit is
\begin{equation}
Z_{m,\perp}^{(s)}[h]\propto
 \left\{\det{}'\!\left[\ell^{2s}A_h^s\right]\right\}^{-D/2},
\label{eq:fixedZdet-long}
\end{equation}
where the prime omits the scalar zero mode and metric-independent normalizations are suppressed. Restoring $x_0$ gives the fixed-metric matter factor in Eq.~\eqref{eq:polyakov-full-long}. In flat space, $A\to |p|^2$, and Eq.~\eqref{eq:fractional-action-long} gives
\begin{equation}
\langle \widetilde X^\mu(p)\widetilde X^\nu(-p)\rangle
=(2\pi\alpha')\ell^{2-2s}g_{\rm E}^{\mu\nu}|p|^{-2s}.
\label{eq:flatprop-long}
\end{equation}
For $1/4<s<1$ its position-space kernel is the convergent oscillatory Bessel integral
\begin{align}
G_s(x)&=\int\frac{\dd^2p}{(2\pi)^2}\frac{\e^{\ii p\cdot x}}{|p|^{2s}}
=\frac{1}{2\pi}\int_0^\infty \dd p\,p^{1-2s}J_0(p|x|)  =\frac{\Gamma(1-s)}{4^s\pi\Gamma(s)}|x|^{2s-2} .
\label{eq:flatgreen-long}
\end{align}
The correlator is $2\pi\alpha'\ell^{2-2s}G_s(x)$. The gamma-function expression extends Eq.~\eqref{eq:flatgreen-long} distributionally to $0<s\leq1/4$. As $s\to1$, its finite part after the usual infrared subtraction becomes the logarithmic Green function of the two-dimensional massless scalar.

Equation~\eqref{eq:flatgreen-long} assigns the spectral coordinate the scaling dimension $\Delta_X=1-s$, the worldsheet analogue of the noncanonical dimensions carried by fractional gauge fields and holographic boundary currents~\cite{LaNaveRMP2019,LaNavePhillips2019,HeydemanNonlocalQED2020}.
Worldsheet gravity reads this anomalous dimension a second time: the same exponent becomes a determinant weight and is fixed by Weyl cancellation together with BRST consistency.

For $0<s<1$, the spectral power also has the Balakrishnan representation~\cite{Balakrishnan1960}
\begin{equation}
A^s f=\frac{1}{\Gamma(-s)}\int_0^\infty \dd t\,t^{-1-s}\left(\e^{-tA}-1\right)f,
\label{eq:balak-long}
\end{equation}
on the nonzero-mode subspace. On an eigenmode it reduces to
\begin{equation}
\frac{1}{\Gamma(-s)}\int_0^\infty \dd t\,t^{-1-s}(\e^{-t\lambda}-1)=\lambda^s,
\end{equation}
with the subtraction cancelling the small-$t$ singularity. The same operator also admits the Caffarelli--Silvestre Dirichlet-to-Neumann realization on flat space~\cite{CS2007,FrassinoPanella2019}: if $U(x,y)$ solves
\begin{equation}
\nabla_A\left(y^{1-2s}\nabla^A U\right)=0,
\qquad U(x,0)=X(x),
\end{equation}
then
\begin{align}
(-\Delta)^sX(x)&=-d_s\lim_{y\to0^+}y^{1-2s}\partial_y U(x,y),\nonumber\\
d_s&=\frac{2^{2s-1}\Gamma(s)}{\Gamma(1-s)} .
\label{eq:CS-long}
\end{align}
Appendix~\ref{app:localization} gives the Bessel-function normalization.

In this extension regime, the fractional worldsheet action is the boundary response of one additional weighted coordinate, mirroring the bulk-to-boundary origin of fractional gauge dynamics~\cite{CS2007,LaNavePhillips2019,FrassinoPanella2019}. The compact construction itself is the intrinsic spectral theory defined above for $s>0$.

\section{Spectral determinants and the metric-dependent Jacobian}\label{sec:detjac}

For a positive self-adjoint operator $A$ with nonzero eigenvalues $\lambda_n$, the zeta function and determinant are
\begin{align}
\zeta_A(z)&=\Tr{}'(A^{-z})=\sum_{n\ge1}\lambda_n^{-z},\nonumber\\
\log\det{}'(\ell^2A)&=-\left.\frac{\dd}{\dd z}\big(\ell^{-2z}\zeta_A(z)\big)\right|_{z=0}.
\label{eq:zetadet-long}
\end{align}
The length $\ell$ records the local scale dependence of the determinant line. Since the spectral power uses the same eigenbasis,
\begin{equation}
\zeta_{A^s}(z)=\sum_{n\ge1}(\lambda_n^s)^{-z}=\zeta_A(sz).
\end{equation}
With $w=sz$, its derivative at the origin is
\begin{align}
\log\det{}'(\ell^{2s}A^s)
&=-\left.\frac{\dd}{\dd z}
\left[\ell^{-2sz}\zeta_A(sz)\right]\right|_{z=0} =-s\left.\frac{\dd}{\dd w}
\left[\ell^{-2w}\zeta_A(w)\right]\right|_{w=0}.
\end{align}
Thus
\begin{equation}
\det{}'(\ell^{2s}A^s)=\left[\det{}'(\ell^2A)\right]^s .
\label{eq:detidentity-long}
\end{equation}
The fractional kinetic operator therefore multiplies the scalar determinant exponent by $s$.

Consequently the nonzero-mode fractional determinant is
\begin{equation}
Z_{m,\perp}^{(s)}[h]\propto
 \left[\det{}'(\ell^2A_h)\right]^{-sD/2}.
\label{eq:Zm-sD-long}
\end{equation}
The corresponding map is
\begin{equation}
\mc I_h=P_0(h)+\ell^{s-1}A_h^{(s-1)/2}P_\perp(h),
\qquad Y=\mc I_hX,
\label{eq:Ymap-long}
\end{equation}
with inverse
\begin{equation}
\mc I_h^{-1}=P_0(h)+\ell^{1-s}A_h^{-(s-1)/2}P_\perp(h).
\label{eq:Iinverse-long}
\end{equation}
With their natural operator domains, Eqs.~\eqref{eq:Ymap-long} and \eqref{eq:Iinverse-long} map the quadratic-form spaces
\begin{equation}
\mathbb R\oplus\bigl(H^s(\Sigma)\cap\Ran P_\perp\bigr)
\ \longleftrightarrow\
\mathbb R\oplus\bigl(H^1(\Sigma)\cap\Ran P_\perp\bigr).
\label{eq:Sobolev-isomorphism-long}
\end{equation}
The maps are inverse to one another. They fix $P_0Y=P_0X=x_0$, while for $n\ge1$,
$Y_n=(\ell^2\lambda_n)^{(s-1)/2}X_n$. Mode by mode,
\begin{equation}
\ell^{2s-2}\lambda_n^s |X_n|^2=\lambda_n |Y_n|^2,
\qquad n\ge1.
\end{equation}
Summing over modes gives
\begin{equation}
S_s[X;h]=S_1[Y;h]
\label{eq:actionintertwine-long}
\end{equation}
for each fixed $h$.

At a finite spectral cutoff $M$, expand each target component as
\begin{equation}
X^\mu=X_0^\mu\varphi_0+\sum_{n=1}^M X_n^\mu\varphi_n,
\qquad X_0^\mu=\Area[h]^{1/2}x_0^\mu,
\end{equation}
and define the $L^2(h)$-ultralocal mode measure by
\begin{equation}
\mc D_h^{(M)}X=\prod_{\mu=1}^D\prod_{n=0}^M\dd X_n^\mu .
\label{eq:Xmeasure-long}
\end{equation}
The diagonal change of variables $X_n\mapsto Y_n$ gives
\begin{equation}
\mc D_h^{(M)}X=
 \prod_{n=1}^M(\ell^2\lambda_n)^{-(s-1)D/2}\mc D_h^{(M)}Y.
\end{equation}
Its zeta-regularized limit is
\begin{equation}
\mc D_hX=
 \left[\det{}'(\ell^2A_h)\right]^{-(s-1)D/2}\mc D_hY .
\label{eq:jac-long}
\end{equation}
Since $\mc I_h$ fixes $x_0$, the target-coordinate measure contributes the separate factor $\Area[h]^{D/2}$ used in Sec.~\ref{sec:anomaly}.

At fixed $h$, Eqs.~\eqref{eq:actionintertwine-long} and \eqref{eq:jac-long} are the Hilbert-space localization of
Ref.~\cite{BasaClassification}. Integrating over $h$ changes its role: $\mc I_h$ now varies over metric space, and its Jacobian becomes the moduli density and Weyl anomaly. The fixed-background equivalence thereby acquires quantum-geometric content.

\section{Weyl anomaly of the spectral matter measure}\label{sec:anomaly}

Let $h(t)=e^{2t\sigma}h$ and let $\lambda_n(t)$ be a nonzero scalar eigenvalue normalized in $L^2(h(t))$. The Dirichlet form
\begin{equation}
q_h[u,v]=\int_\Sigma\dd\mu_h\,h^{ab}\partial_a u\,\partial_b v
\label{eq:Dirichlet-form-long}
\end{equation}
is Weyl invariant in two dimensions, whereas
\begin{equation}
\left.\frac{\dd}{\dd t}\langle u,v\rangle_{h(t)}\right|_{t=0}
=2\int_\Sigma\dd\mu_h\,\sigma\,uv .
\label{eq:L2-Weyl-long}
\end{equation}
Differentiating the weak eigenvalue equation
$q_h[\varphi_n,v]=\lambda_n\langle\varphi_n,v\rangle_h$
and setting $v=\varphi_n$ cancels the variation of $\varphi_n$ and leaves
\begin{equation}
\dot\lambda_n(0)
=-2\lambda_n\int_\Sigma\dd\mu_h\,\sigma\,\varphi_n^2 .
\label{eq:eigenvalue-Weyl-long}
\end{equation}
For a degenerate eigenvalue, tracing over its eigenspace gives the same variation and directly determines the determinant response.

With a proper-time cutoff, the metric-dependent part of the determinant is
\begin{equation}
\log\det{}'_\epsilon(\ell^2A_h)
=-\int_\epsilon^\infty\frac{\dd\tau}{\tau}
\sum_{n\ge1}\e^{-\tau\lambda_n}
+2\log\ell\,\zeta_{A_h}(0).
\label{eq:proper-time-det-long}
\end{equation}
Since $\zeta_{A_h}(0)$ is topological on a closed surface, its Weyl variation vanishes. Varying the eigenvalues in Eq.~\eqref{eq:proper-time-det-long} and using Eq.~\eqref{eq:eigenvalue-Weyl-long} gives
\begin{align}
\delta_\sigma\log\det{}'_\epsilon A_h
&=\sum_{n\ge1}\int_\epsilon^\infty\dd\tau\,
\delta_\sigma\lambda_n\,\e^{-\tau\lambda_n} =-2\sum_{n\ge1}\e^{-\epsilon\lambda_n}
\int_\Sigma\dd\mu_h\,\sigma\,\varphi_n^2\nonumber\\
&=-2\,\Tr{}'\!\left(\sigma\e^{-\epsilon A_h}\right).
\label{eq:prime-heat-variation-long}
\end{align}
The diagonal scalar heat kernel is
\begin{equation}
K(\epsilon;x,x)
=\frac{1}{4\pi\epsilon}
+\frac{R[h](x)}{24\pi}
+O(\epsilon).
\label{eq:heat-diagonal-long}
\end{equation}
Appendix~\ref{app:heatkernel} derives the coefficient $R/6$ from the heat-kernel transport equation. Removing the normalized constant mode gives
\begin{align}
\Tr{}'\!\left(\sigma\e^{-\epsilon A_h}\right)
&=\frac{1}{4\pi\epsilon}\int_\Sigma\dd\mu_h\,\sigma
+\frac{1}{24\pi}\int_\Sigma\dd\mu_h\,\sigma R -\frac{1}{\Area[h]}\int_\Sigma\dd\mu_h\,\sigma
+O(\epsilon).
\label{eq:prime-heat-expansion-long}
\end{align}
After renormalizing the local cosmological divergence,
\begin{align}
\delta_\sigma\log\det{}'(\ell^2A_h)
&=-\frac{1}{12\pi}\int_\Sigma\dd\mu_h\,\sigma R[h]
 +\frac{2}{\Area[h]}\int_\Sigma\dd\mu_h\,\sigma .
\label{eq:det-Weyl-inf-long}
\end{align}
The final term is $\delta_\sigma\log\Area[h]$.

For $h=e^{2\varphi}\hat h$ one has
\begin{equation}
\dd\mu_h R[h]
=\dd\mu_{\hat h}\left(\hat R-2\hat\Delta\varphi\right).
\label{eq:curvature-Weyl-long}
\end{equation}
The Liouville functional
\begin{equation}
S_L[\varphi;\hat h]=\int_\Sigma\dd\mu_{\hat h}
\left[(\hat\nabla\varphi)^2+\hat R\varphi\right]
\label{eq:SL-long}
\end{equation}
satisfies
\begin{equation}
\delta S_L[\varphi;\hat h]
=\int_\Sigma\dd\mu_h\,\delta\varphi\,R[h].
\label{eq:SL-variation-long}
\end{equation}
Integrating Eq.~\eqref{eq:det-Weyl-inf-long} along the conformal path gives the closed-surface Polyakov--Alvarez relation~\cite{Polyakov1981,Alvarez1983,Vassilevich2003},
\begin{align}
\log\frac{\det{}'(\ell^2A_{e^{2\varphi}\hat h})}
 {\det{}'(\ell^2A_{\hat h})}
&=-\frac{1}{12\pi}S_L[\varphi;\hat h]  +\log\frac{\Area[e^{2\varphi}\hat h]}{\Area[\hat h]} .
\label{eq:PA-long}
\end{align}
Dividing by the dimensionless area defines
\begin{equation}
\mathfrak D[h]=\frac{\det{}'(\ell^2A_h)}{\Area[h]/\ell^2}
\label{eq:area-det-long}
\end{equation}
and removes the zero-mode term:
\begin{equation}
\log\frac{\mathfrak D[e^{2\varphi}\hat h]}{\mathfrak D[\hat h]}
=-\frac{1}{12\pi}S_L[\varphi;\hat h].
\label{eq:normalized-PA-long}
\end{equation}

The scalar zero mode contributes $\dd X_0=\Area[h]^{1/2}\dd x_0$ per target coordinate. Dividing by $V_D=\int\dd^Dx_0$ gives
\begin{equation}
\frac{Z_m^{(s)}[h]}{V_D}\propto
 [\Area[h]/\ell^2]^{D/2}
 \left[\det{}'(\ell^2A_h)\right]^{-sD/2}.
\label{eq:full-area-Z-long}
\end{equation}
Combining Eqs.~\eqref{eq:PA-long} and \eqref{eq:full-area-Z-long} gives the finite transformation law
\begin{align}
\log\frac{Z_m^{(s)}[e^{2\varphi}\hat h]}{Z_m^{(s)}[\hat h]}
&=\frac{sD}{24\pi}S_L[\varphi;\hat h] -\frac{(s-1)D}{2}
 \log\frac{\Area[e^{2\varphi}\hat h]}{\Area[\hat h]}.
\label{eq:full-weyl-long}
\end{align}
On the fixed-area slice the second line vanishes. The curvature part of an area-preserving infinitesimal Weyl response is
\begin{equation}
\langle T^a{}_a\rangle_h\big|_R=-\frac{sD}{24\pi}R[h],
 \qquad c_{\rm det}=sD.
\label{eq:trace-anom-long}
\end{equation}
The area constraint fixes the constant trace component; Sec.~\ref{sec:completion} realizes the remaining determinant weight in a local CFT.

Worldsheet gravity converts the scaling exponent into a Weyl weight: each spectral coordinate contributes $s$, rather than one, to the determinant line. Ghost cancellation then turns anomalous scaling into the consistency condition $sD=26$, or $D_\ast=26/s$---a constraint with no fixed-background counterpart.

\section{Polyakov gauge fixing and the moduli determinant density}\label{sec:gaugefix}

The full theory is defined by the Polyakov functional integral
\begin{equation}
\mc Z_s=\frac{1}{\Vol(\mathrm{Diff}(\Sigma)\times\mathrm{Weyl}(\Sigma))}
\int \mc D h\,\mc D X\,\e^{-S_s[X;h]}.
\label{eq:polyakov-full-long}
\end{equation}
The worldsheet group $\mathrm{Diff}(\Sigma)\times\mathrm{Weyl}(\Sigma)$ acts locally on $Y$ and is transported to $X$ by $\mc I_h$. For a diffeomorphism $f$ and Weyl factor $\omega$, let
\begin{equation}
h'=e^{2\omega}f^*h,
\qquad
X'=\mc I_{h'}^{-1}f^*(\mc I_hX).
\label{eq:gauge-X-long}
\end{equation}
Then $Y'=\mc I_{h'}X'=f^*Y$. The conjugated maps compose because adjacent intertwiners cancel. Writing
\begin{equation}
U^{(X)}_{g;h}=\mc I_{h^g}^{-1}U_{g;h}\mc I_h,
\qquad U_{g;h}Y=f^*Y,
\end{equation}
successive transformations obey
\begin{align}
U^{(X)}_{g_2;h^{g_1}}U^{(X)}_{g_1;h}
&=\mc I_{h^{g_2g_1}}^{-1}U_{g_2;h^{g_1}}
\underbrace{\mc I_{h^{g_1}}\mc I_{h^{g_1}}^{-1}}_{1}
U_{g_1;h}\mc I_h\nonumber\\
&=\mc I_{h^{g_2g_1}}^{-1}U_{g_2g_1;h}\mc I_h
=U^{(X)}_{g_2g_1;h}.
\label{eq:gauge-composition-long}
\end{align}
The Weyl-invariant Dirichlet action of $Y$ then gives
\begin{equation}
S_s[X^g;h^g]
=S_1[U_{g;h}(\mc I_hX);h^g]
=S_1[\mc I_hX;h]
=S_s[X;h].
\label{eq:gauge-invariance-long}
\end{equation}

Choose the fixed-area conformal slice
\begin{equation}
h=e^{2\phi}\hat h(m),
\qquad \frac{\Area[h]}{\ell^2}
=\frac{\Area[\hat h(m)]}{\ell^2}=1,
\label{eq:fixed-area-slice-long}
\end{equation}
where the Liouville zero mode is fixed and $m$ are moduli. Faddeev--Popov gauge fixing gives the ghost action and moduli insertions~\cite{David1988,DistlerKawai1989,Polchinski,GuillarmouRhodesVargas2019}.
The global area term vanishes on this slice, and the remaining Jacobian defines the moduli density
\begin{equation}
\mc J_s[h]\equiv\mathfrak D[h]^{-(s-1)D/2},
\qquad
\mc J_s[m]\equiv\mc J_s[\hat h(m)] .
\label{eq:Jsm-long}
\end{equation}
Relative to $D$ ordinary $Y$ bosons, $\mc J_s[m]$ shifts the integrated anomaly coefficient to
\begin{equation}
c_{\rm det}+c_{\gh}=sD-26.
\end{equation}
The gauge-fixed vacuum functional is
\begin{widetext}
\begin{align}
\mc Z_s
&=\frac{1}{\Vol(\mathrm{Weyl}_{\rm FA})}
\int_{\mc M_g}\dd m\,\mc J_s[m]
\int_{\int\sqrt{\hat h}(e^{2\phi}-1)=0}
 \mc D\phi\,\mc D b\,\mc D c\,\mc D Y\;\nonumber\\
&\quad\times
\exp\Big[-S_1[Y;\hat h(m)]-S_{\rm gh}[b,c;\hat h(m)]
+\frac{sD-26}{24\pi}S_L[\phi;\hat h(m)]\Big],
\label{eq:gaugefixed-long}
\end{align}
\end{widetext}
Here $\dd m$ includes the moduli $b$-ghost insertions and the conformal-Killing reduction of  ppendix~\ref{app:gaugefix}; metric-independent normalizations are suppressed. At $sD=26$ the Liouville factor disappears and the fixed-area Weyl quotient cancels. The surviving family $\mathfrak D[\hat h(m)]$ is the Quillen norm of a scalar determinant-line section over moduli space~\cite{Quillen1985}.

\section{Explicit local completion on the critical branch}\label{sec:completion}

\paragraph{Critical-completion theorem.}
Let $D$ be an integer with $1\leq D\leq26$ and impose
\begin{equation}
s=\frac{26}{D},\qquad N=(s-1)D=26-D\in\mathbb Z_{\ge0}.
\label{eq:critical-N-long}
\end{equation}
Introduce $N$ decoupled scalar fields $\chi^I$ with
\begin{equation}
S_\chi=\frac{1}{4\pi\alpha'}\sum_{I=1}^N
 \langle P_\perp\chi^I,A_hP_\perp\chi^I\rangle_h
\label{eq:chi-action-long}
\end{equation}
with the same area-normalized zero-mode convention as $Y$. Their Gaussian integral is
\begin{equation}
Z_\chi[h]=\mathfrak D[h]^{-N/2}=\mc J_s[h].
\label{eq:chi-determinant-long}
\end{equation}
Writing $\vert_{\rm FA}$ for $\Area[h]/\ell^2=1$, this local realization gives the genus-by-genus identity
\begin{align}
\left.\frac{Z_m^{(s)}[h]}{V_D}\right|_{\rm FA}
&=\mathfrak D[h]^{-13} =\left.\frac{Z_Y[h]}{V_D}\right|_{\rm FA}
Z_\chi[h],\qquad sD=26 .
\label{eq:fixed-area-completion-long}
\end{align}
Here $Z_Y$ is the partition function of the $D$ ordinary $Y$ bosons, and the spectator zero-mode volume is divided out of $Z_\chi$. The matter state space is
\begin{align}
\mc H_m&=\mc H_Y\otimes\mc H_\chi,
&T_m&=T_Y+T_\chi,\nonumber\\
c_m&=D+N=26.
\label{eq:completed-cft-long}
\end{align}
The spectator determinant is an operator sector: its states run through plumbing channels and complete the local stress tensor. The $X$ observables remain spectral composites of $Y$, so their anomalous scaling survives inside a unitary factorizing CFT. This free-field completion exists for integral $N=26-D\ge0$; other values retain the spectral determinant measure without a finite spectator-boson realization.

\section{Genus-one partition function, sewing, and BRST structure}\label{sec:genusbrst}

\subsection{Torus determinant and modular invariance}

On the unit-area flat torus with coordinates
$0\leq\sigma^1,\sigma^2<1$ and metric
\begin{equation}
\dd s^2=\frac{\left|\dd\sigma^1+\tau\,\dd\sigma^2\right|^2}{\tau_2},
\end{equation}
the nonzero scalar eigenvalues are
\begin{equation}
\lambda_{m,n}=\frac{4\pi^2}{\tau_2}|m\tau-n|^2,
\qquad (m,n)\in\mathbb Z^2\setminus\{(0,0)\}.
\label{eq:torus-eigenvalues-long}
\end{equation}
The zeta-regularized lattice product in Appendix~\ref{app:torus} gives, up to a modulus-independent constant,
\begin{equation}
\mathfrak D(\tau)\propto \tau_2|\eta(\tau)|^4 .
\end{equation}
Raising it to the fractional matter exponent gives
\begin{equation}
Z_m^{(s)}(\tau)\propto \left(\tau_2|\eta(\tau)|^4\right)^{-sD/2}.
\label{eq:torusdet-long}
\end{equation}
In oscillator form,
\begin{align}
Z_m^{(s)}(\tau)&\propto \tau_2^{-sD/2}|q|^{-sD/12}
 \prod_{n=1}^\infty|1-q^n|^{-2sD},\nonumber\\
q&=\e^{2\pi\ii\tau}.
\label{eq:torus-product-long}
\end{align}
The $D$ fields $Y$ carry the power $-D/2$, and the $N=(s-1)D$ spectators carry
\begin{equation}
Z_\chi(\tau)=\mc J_s(\tau)\propto
 \left(\tau_2|\eta(\tau)|^4\right)^{-N/2}.
\end{equation}
Since $\tau_2|\eta(\tau)|^4$ is invariant under $SL(2,\mathbb Z)$, the vacuum determinant is modular invariant; Eq.~\eqref{eq:torus-product-long} separates its Casimir factor from its oscillators.

\subsection{Degeneration and sewing}

In the torus degeneration $q\to0$,
\begin{equation}
Z_m^{(s)}(\tau)\sim
\left[-\frac{1}{2\pi}\log|q|\right]^{-sD/2}|q|^{-sD/12},
\label{eq:torus-degeneration-long}
\end{equation}
up to a constant and oscillator corrections. The power of $|q|$ fixes the Casimir energy, while the plumbing cylinder inserts the identity on
\begin{equation}
\mc H_m=\mc H_Y\otimes\mc H_\chi.
\end{equation}
Let $\{|a\rangle\}$ diagonalize $L_0$ and $\bar L_0$ on $\mc H_m$, with BPZ metric $G_{ab}=\langle a|b\rangle$ and integration over continuous momenta. For two surfaces joined by a cylinder, the matter contribution is
\begin{equation}
\sum_{a,b\in\mc H_m}
 \vev{\mc O_L|a}(G^{-1})^{ab}
 \left\langle b\left|
 q^{L_0-c_m/24}\bar q^{\bar L_0-c_m/24}
 \right|\mc O_R\right\rangle ,
\label{eq:sewing-long}
\end{equation}
where the sums include the corresponding integrals, ghost insertion, and BRST projection~\cite{BPZ1984,Polchinski}. Its trace is Eq.~\eqref{eq:torus-product-long}. External operators $\mc O_Y\otimes\mathbf1_\chi$ carry the spectator vacuum, but the internal channel runs over the tensor-product spectrum.

\subsection{BRST charge in the local completion}

The completed matter stress tensor is
\begin{align}
T_m(z)&=-\frac1{\alpha'}:\!\partial Y^\mu\partial Y_\mu\!:
-\frac1{\alpha'}:\!\partial\chi^I\partial\chi_I\!:,\nonumber\\
c_m&=D+N=26.
\label{eq:Tm-completion-long}
\end{align}
Together with
\begin{equation}
T_{\gh}(z)=-2:\!b\partial c\!:-:\!(\partial b)c\!:,
\qquad c_{\gh}=-26,
\end{equation}
the holomorphic BRST current and charge are
\begin{equation}
 j_B=c\left(T_m+\frac12T_{\gh}\right)+\frac32\partial^2c,
 \qquad Q_B=\oint\frac{\dd z}{2\pi\ii}\,j_B(z),
\label{eq:BRST-current-long}
\end{equation}
with the antiholomorphic copy. Wick contractions leave the obstruction
\begin{equation}
Q_B^2\ \propto\ (c_m-26)\oint\frac{\dd z}{2\pi\ii}\,
 c\,\partial^3c,
\label{eq:Q2-long}
\end{equation}
up to the normalization of $Q_B$~\cite{KatoOgawa1983,Polchinski}. With the usual intercept, $c_m=26$ gives
\begin{equation}
Q_B^2=\bar Q_B^2=\{Q_B,\bar Q_B\}=0 .
\label{eq:criticality-long}
\end{equation}
The torus trace, plumbing identity, and nilpotent BRST charge are three views of the same local completion: they govern the vacuum, intermediate states, and physical cohomology, respectively. The spectral sector enters through its Polyakov measure, while the $Y+\chi$ Virasoro theory furnishes the operator algebra.

\section{Observables and target-space backgrounds}\label{sec:observables}

\subsection{Zero modes and vertex operators}

Because $\mc I_h$ fixes the translation mode,
\begin{equation}
Y^\mu=x_0^\mu+\ell^{s-1}A_h^{(s-1)/2}\widetilde X^\mu .
\label{eq:Yzero-long}
\end{equation}
The $x_0$ integration gives the target volume in the vacuum functional and momentum conservation in amplitudes:
\begin{equation}
\int_{\mathbb R^D}\dd^Dx_0\,
\exp\!\left(\ii x_0\cdot\sum_i k_i\right)
=(2\pi)^D\delta^{(D)}\!\left(\sum_i k_i\right).
\label{eq:zero-mode-delta-long}
\end{equation}
For the oscillators,
\begin{equation}
\langle\widetilde Y^\mu(z)\widetilde Y^\nu(w)\rangle
=-\frac{\alpha'}{2}g_{\rm E}^{\mu\nu}\log|z-w|^2,
\label{eq:Y-propagator-long}
\end{equation}
normal ordering and Wick contraction give
\begin{widetext}
\begin{align}
\left\langle
\exp\!\left(\ii\sum_i k_i\cdot\widetilde Y(z_i)\right)
\right\rangle
&=\exp\!\left[
-\sum_{i<j}k_i\cdot k_j
\langle\widetilde Y(z_i)\widetilde Y(z_j)\rangle
\right]\nonumber\\
&=\prod_{i<j}|z_i-z_j|^{\alpha' k_i\cdot k_j}.
\label{eq:Gaussian-KN-long}
\end{align}
Thus exponential vertices in the spectator-identity subsector obey
\begin{equation}
\left\langle\prod_{i=1}^{n}\left({:}\e^{\ii k_i\cdot Y(z_i,\bar z_i)}{:}
 \otimes\mathbf1_\chi\right)\right\rangle
=(2\pi)^D\delta^{(D)}\!\left(\sum_i k_i\right)
\prod_{i<j}|z_i-z_j|^{\alpha' k_i\cdot k_j},
\label{eq:KNcorrelator-long}
\end{equation}
\end{widetext}
The unintegrated closed-string tachyon vertex is
\begin{equation}
V_T=c\bar c\,{:}\e^{\ii k\cdot Y}{:}\otimes\mathbf1_\chi,
\qquad
(h,\bar h)=\left(\frac{\alpha'k^2}{4},\frac{\alpha'k^2}{4}\right),
\label{eq:tachyon-vertex-long}
\end{equation}
so BRST closure gives $k^2=4/\alpha'$ in the Euclidean momentum convention. Fixing three punctures at $0$, $1$, and $\infty$, the four-tachyon matter block becomes
\begin{widetext}
\begin{equation}
\mc A_4^{(s)}
=\mc N_4(2\pi)^D\delta^{(D)}\!\left(\sum_{i=1}^4k_i\right)
\int_{\mathbb C}\dd^2z\,
|z|^{\alpha'k_1\cdot k_2}
|1-z|^{\alpha'k_2\cdot k_3},
\label{eq:four-tachyon-long}
\end{equation}
\end{widetext}
with analytic continuation to Lorentzian Mandelstam invariants~\cite{KobaNielsen1969}. In the spectral variable, the exponential vertex is
\begin{equation}
{:}\e^{\ii k\cdot Y(z)}{:}=
\e^{\ii k\cdot x_0}
{:}\exp\!\left[\ii k\cdot\ell^{s-1}
 (A_h^{(s-1)/2}\widetilde X)(z)\right]{:}_Y .
\label{eq:vertex-pullback-long}
\end{equation}
Normal ordering is defined by Eq.~\eqref{eq:Y-propagator-long}; the first factor carries the conserved target momentum.

The massless vertex in this distinguished subsector is
\begin{equation}
\mc V_\varepsilon=\frac{1}{2\pi\alpha'}\int\dd^2z\,
\varepsilon_{\mu\nu}\,{:}\partial Y^\mu\bar\partial Y^\nu
 \e^{\ii k\cdot Y}{:}\otimes\mathbf1_\chi,
\label{eq:masslessvertex-long}
\end{equation}
with BRST constraints
\begin{equation}
k^2=0,
\qquad
k^\mu\varepsilon_{\mu\nu}=0,
\qquad
k^\nu\varepsilon_{\mu\nu}=0,
\end{equation}
and the gauge equivalence
\begin{equation}
\varepsilon_{\mu\nu}\sim\varepsilon_{\mu\nu}
+k_\mu\xi_\nu+\widetilde\xi_\mu k_\nu .
\end{equation}
The symmetric, antisymmetric, and trace components are the metric, Kalb--Ramond, and dilaton perturbations of the $Y$ sector. The completed string spectrum also contains states generated by $\chi$ oscillators and momenta.

The distinguished massless states are ordinary BRST cohomology classes in $Y$ and nonlocal spectral composites in $X$. Their polarization conditions are unchanged, while $s$ remains visible in the map between the two representations. Anomalous scaling therefore lives inside a factorizing closed-string sector, paralleling fractional gauge theories with noncanonical field dimensions~\cite{LaNaveRMP2019,LaNavePhillips2019}.

\subsection{Target-space background completion}

Target backgrounds couple to the intrinsic map $Y:\Sigma\to\mc M$, with the spectator CFT free:
\begin{widetext}
\begin{align}
S_\sigma[Y;G,B,\Phi]
&=\frac{1}{4\pi\alpha'}\int\dd^2\sigma\sqrt h\,
\left(h^{ab}G_{\mu\nu}(Y)+\ii\epsilon^{ab}B_{\mu\nu}(Y)\right)
\partial_aY^\mu\partial_bY^\nu
\nonumber\\
&\qquad +\frac{1}{4\pi}\int\dd^2\sigma\sqrt h\,R^{(2)}\Phi(Y).
\label{eq:sigma-action-long}
\end{align}
\begin{align}
Z_s[h;G,B,\Phi]
&:=\int\mc D_hX\,
\e^{-S_\sigma[\mc I_hX;G,B,\Phi]}
\nonumber\\
&=\left[\det{}'(\ell^2A_h)\right]^{-(s-1)D/2}
\int\mc D_hY\,\e^{-S_\sigma[Y;G,B,\Phi]} .
\label{eq:interacting-pullback-long}
\end{align}
\end{widetext}
The inverse intertwiner defines $X=\mc I_h^{-1}Y$ chartwise on the
mapping space.
Here $\epsilon^{ab}=\varepsilon^{ab}/\sqrt h$ is the Levi--Civita tensor with $\varepsilon^{12}=1$, and $H_{\mu\nu\rho}=3\partial_{[\mu}B_{\nu\rho]}$.
Equation~\eqref{eq:Yzero-long} then gives the pseudodifferential sigma
model in spectral $X$ coordinates.

The background-field expansion in Appendix~\ref{app:bg} gives the one-loop Weyl-anomaly coefficients~\cite{Friedan1980,Callan1985,FradkinTseytlin1985}:
\begin{widetext}
\begin{align}
\bar\beta^G_{\mu\nu}&=\alpha'\left(R_{\mu\nu}-\frac14 H_{\mu\rho\sigma}H_\nu{}^{\rho\sigma}+2\nabla_\mu\nabla_\nu\Phi\right)+O(\alpha'^2),\nonumber\\
\bar\beta^B_{\mu\nu}&=\alpha'\left(-\frac12\nabla^\rho H_{\rho\mu\nu}+\nabla^\rho\Phi\,H_{\rho\mu\nu}\right)+O(\alpha'^2).
\label{eq:GBbeta-long}
\end{align}
\end{widetext}
The spectral determinant shifts the scalar Weyl term, giving
\begin{align}
\bar\beta^\Phi&=\frac{sD-26}{6}
+\alpha'\Big[(\nabla\Phi)^2-\frac12\nabla^2\Phi
-\frac{1}{24}H_{\mu\nu\rho}H^{\mu\nu\rho}\Big]
+O(\alpha'^2).
\label{eq:phibeta-long}
\end{align}
At $sD=D+N=26$, $B=0$ and constant dilaton reduce Eq.~\eqref{eq:GBbeta-long} to
\begin{equation}
R_{\mu\nu}=0.
\end{equation}
For flat $G_{\mu\nu}$, $B=0$, and $\Phi=V_\mu Y^\mu$, Eq.~\eqref{eq:phibeta-long} gives
\begin{equation}
sD+6\alpha' V^2=26.
\label{eq:lineardilaton-long}
\end{equation}

These beta functions are the Euler--Lagrange equations of the string-frame action
\begin{widetext}
\begin{equation}
S_{\rm eff}
=\frac{1}{2\kappa^2}\int\dd^DY\,\sqrt G\,\e^{-2\Phi}
\left[
R+4(\nabla\Phi)^2-\frac1{12}H_{\mu\nu\rho}H^{\mu\nu\rho}
-\frac{2(sD-26)}{3\alpha'}+O(\alpha')
\right].
\label{eq:target-effective-action-long}
\end{equation}
\end{widetext}
The $B_{\mu\nu}$ variation gives
\begin{equation}
\nabla^\rho\!\left(\e^{-2\Phi}H_{\rho\mu\nu}\right)=0,
\end{equation}
and the metric equation combines with the dilaton variation
\begin{equation}
R+4\nabla^2\Phi-4(\nabla\Phi)^2
-\frac1{12}H^2-\frac{2(sD-26)}{3\alpha'}=0.
\label{eq:dilaton-EOM-long}
\end{equation}
For flat $G$, $H=0$, and $\Phi=V\cdot Y$,
Eq.~\eqref{eq:dilaton-EOM-long} becomes Eq.~\eqref{eq:lineardilaton-long}. Higher orders use the local curvature--$H$--dilaton operator basis with the shifted deficit.

The spectral deformation is global in the local representative: it changes the determinant weight and dilaton balance, while the one-loop tensor geometry remains that of the $Y$ sigma model. This is the target-space image of the separation between anomalous scaling and local covariance on the worldsheet~\cite{LaNaveRMP2019,LaNavePhillips2019}.

\section{Conclusions}

We have shown that a fractional worldsheet exponent enters quantum gravity through the measure, not only through the fixed-metric propagator. The metric-dependent intertwiner maps the action to free bosons, while its Jacobian changes the determinant-line weight to $sD$. Weyl cancellation therefore selects $sD=26$, or
$D_\ast=26/s$.

On the integer critical locus, $26-D$ spectator bosons turn this determinant into a local $c_m=26$ CFT. Modular invariance, plumbing, and BRST cohomology then arise from one tensor-product Hilbert space, while the distinguished $X$ coordinates remain nonlocal spectral composites. Their pullback carries the Koba--Nielsen and massless
sectors and couples to target backgrounds with dilaton deficit $sD-26$.

The same exponent that measures nonlocal response in quantum matter thus becomes a consistency parameter of fluctuating two-dimensional geometry. Fixed-metric localization is only the first step: as the localization map moves over metric space, its Jacobian records the fractional scaling in the quantum geometry of the worldsheet.

\appendix

\section{Localization constants and flat-space integrals}\label{app:localization}

\subsection{Balakrishnan semigroup formula}

For $\lambda>0$ and $0<s<1$, set
\begin{equation}
I_s(\lambda)=\int_0^\infty\dd t\,t^{-1-s}
\left(\e^{-\lambda t}-1\right).
\end{equation}
The endpoint behavior removes the boundary term in integration by parts with $v=-t^{-s}/s$:
\begin{align}
I_s(\lambda)
=-\frac{\lambda}{s}\int_0^\infty\dd t\,t^{-s}\e^{-\lambda t}
=-\frac{\lambda^s}{s}\Gamma(1-s)
=\lambda^s\Gamma(-s).
\end{align}
Hence
\begin{equation}
\lambda^s=\frac{1}{\Gamma(-s)}
\int_0^\infty\dd t\,t^{-1-s}
\left(\e^{-\lambda t}-1\right).
\end{equation}
Functional calculus applies the scalar identity mode by mode and gives Eq.~\eqref{eq:balak-long}. Its resolvent form follows from
\begin{equation}
A(A+t)^{-1}=\int_0^\infty\dd u\,A\e^{-u(A+t)}
\end{equation}
and the Euler reflection identity
$\Gamma(s)\Gamma(1-s)=\pi/\sin\pi s$.

\subsection{Caffarelli--Silvestre normalization}

Fourier transformation of $U(p,y)=\widehat X(p)\varphi_s(|p|y)$ fixes the Caffarelli--Silvestre normalization. The extension equation
\begin{equation}
\nabla_A(y^{1-2s}\nabla^A U)=0
\end{equation}
reduces to
\begin{equation}
\varphi_s''(t)+\frac{1-2s}{t}\varphi_s'(t)-\varphi_s(t)=0,
\qquad
\varphi_s(0)=1.
\end{equation}
The decaying solution is
\begin{equation}
\varphi_s(t)=\frac{2^{1-s}}{\Gamma(s)}t^sK_s(t).
\end{equation}
Using
\begin{equation}
K_s(t)\sim 2^{s-1}\Gamma(s)t^{-s}+2^{-s-1}\Gamma(-s)t^s+\cdots
\end{equation}
gives
\begin{align}
\varphi_s(t)&=1+\frac{\Gamma(-s)}{\Gamma(s)}\left(\frac{t}{2}\right)^{2s}+\cdots,\nonumber\\
\varphi_s'(t)&\sim \frac{\Gamma(-s)}{\Gamma(s)}s2^{1-2s}t^{2s-1}.
\end{align}
Thus
\begin{equation}
\lim_{y\to0^+}y^{1-2s}\partial_yU(p,y)
=\widehat X(p)|p|^{2s}\frac{\Gamma(-s)}{\Gamma(s)}s2^{1-2s},
\end{equation}
and, using $\Gamma(1-s)=-s\Gamma(-s)$,
\begin{equation}
|p|^{2s}\widehat X(p)=-\frac{2^{2s-1}\Gamma(s)}{\Gamma(1-s)}
\lim_{y\to0^+}y^{1-2s}\partial_yU(p,y).
\end{equation}
This is Eq.~\eqref{eq:CS-long}.

\subsection{Bessel transform and convergence domain}

The Bessel integral in Eq.~\eqref{eq:flatgreen-long} uses
\begin{equation}
\int_0^\infty \dd p\,p^{\nu-1}J_0(pr)=\frac{2^{\nu-1}\Gamma(\nu/2)}{r^\nu\Gamma(1-\nu/2)},
\qquad 0<\Re\nu<\frac32,
\end{equation}
as an oscillatory integral. Setting $\nu=2-2s$ gives Eq.~\eqref{eq:flatgreen-long} for $1/4<s<1$ and its gamma-function continuation for $0<s\leq1/4$.

\subsection{Fixed-metric spectral positivity}

For $0<s<1$ the fractional propagator admits the positive
K\"all\'en--Lehmann representation
\begin{equation}
(p^2)^{-s}=\frac{\sin(\pi s)}{\pi}
 \int_0^\infty\frac{\dd M^2\,(M^2)^{-s}}{p^2+M^2}.
\label{eq:KL-long}
\end{equation}
Setting $M^2=p^2u$ evaluates the spectral integral through
\begin{equation}
\int_0^\infty\frac{\dd u\,u^{-s}}{1+u}
=B(1-s,s)
=\Gamma(1-s)\Gamma(s)
=\frac{\pi}{\sin\pi s}.
\end{equation}
The spectral density
$\rho_s(M^2)=\sin(\pi s)(M^2)^{-s}/\pi$ is positive throughout this
range.

\section{Heat-kernel coefficient, area term, and determinant scale}\label{app:heatkernel}

For the positive scalar Laplacian $A=-\nabla^2$, let
$\mathcal S(x,y)=\tfrac12d(x,y)^2$ and
$\Delta_{\rm VM}(x,y)$ be the Van Vleck determinant. The local heat-kernel parametrix is
\begin{align}
K(t;x,y)
&\sim\frac{\e^{-\mathcal S(x,y)/(2t)}}{(4\pi t)^{d/2}}
\Delta_{\rm VM}^{1/2}(x,y)\nonumber\\
&\quad\times\sum_{r=0}^\infty t^r u_r(x,y),
\qquad u_0=1.
\end{align}
Substitution into $(\partial_t-\nabla_x^2)K=0$ gives the transport recursion
\begin{equation}
\left(\mathcal S^{;a}\nabla_a+r\right)u_r
=\Delta_{\rm VM}^{-1/2}\nabla^2
\left(\Delta_{\rm VM}^{1/2}u_{r-1}\right).
\end{equation}
In Riemann normal coordinates $\xi^a$ centered at $y$,
\begin{equation}
\Delta_{\rm VM}^{1/2}(x,y)
=1+\frac1{12}R_{ab}(y)\xi^a\xi^b+O(|\xi|^3).
\end{equation}
The coincidence limit of the $r=1$ equation yields
\begin{equation}
[u_1]
=\left[\Delta_{\rm VM}^{-1/2}
\nabla^2\Delta_{\rm VM}^{1/2}\right]
=\frac16R .
\end{equation}
This gives Eq.~\eqref{eq:heat-diagonal-long} and the global prime trace
\begin{align}
\Tr{}'\e^{-tA_h}
&=\frac{\Area[h]}{4\pi t}
+\frac{1}{24\pi}\int_\Sigma\dd\mu_h\,R-1+O(t)\nonumber\\
&=\frac{\Area[h]}{4\pi t}
+\left(\frac{\chi(\Sigma)}6-1\right)+O(t),
\label{eq:global-prime-heat-long}
\end{align}
where Gauss--Bonnet was used in the second line. The Mellin representation
\begin{equation}
\zeta_{A_h}(z)=\frac1{\Gamma(z)}
\int_0^\infty\dd t\,t^{z-1}\Tr{}'\e^{-tA_h}
\end{equation}
is split at $t=1$ and analytically continued after subtracting the displayed small-$t$ terms. Since $1/\Gamma(z)=z+O(z^2)$, the $t^0$ coefficient gives
\begin{equation}
\zeta_{A_h}(0)=\frac{\chi(\Sigma)}6-1.
\label{eq:zeta-zero-long}
\end{equation}

Omitting the zero mode produces the area factor in the prime determinant:
\begin{equation}
\log\frac{\det{}'(\ell^2A_{e^{2\varphi}\hat h})/
[\Area(e^{2\varphi}\hat h)/\ell^2]}
{\det{}'(\ell^2A_{\hat h})/[\Area(\hat h)/\ell^2]}
=-\frac{1}{12\pi}S_L[\varphi;\hat h].
\end{equation}
Restricting to Eq.~\eqref{eq:fixed-area-slice-long} sets the global area ratio to unity and leaves the Liouville term.

A change of reference length in Eq.~\eqref{eq:zetadet-long} shifts
\begin{equation}
\log\det{}'(\ell^2A)\mapsto
\log\det{}'(\ell^2A)+2\log(\ell'/\ell)\zeta_A(0),
\end{equation}
which is a local topological term in two dimensions. For a constant Weyl factor $\varphi=c$, spectral scaling gives
\begin{equation}
\log\frac{\det{}'A_{\e^{2c}h}}{\det{}'A_h}
=-2c\,\zeta_{A_h}(0)
=2c-\frac{\chi(\Sigma)}{3}c .
\end{equation}
This is the constant-$\varphi$ specialization of the Polyakov--Alvarez law: the curvature term is $-\chi c/3$ and the area term is $2c$. The reference-length shift changes only the local topological coefficient, leaving $c_{\rm det}=sD$.

\section{Conformal-gauge Faddeev--Popov reduction}\label{app:gaugefix}

The DeWitt inner product decomposes metric variations into Weyl, diffeomorphism, and moduli directions:
\begin{align}
(\delta h,\delta h)_h
&=\int_\Sigma\dd\mu_h\,
\delta h_{ab}\delta h_{cd} \left(h^{ac}h^{bd}+C\,h^{ab}h^{cd}\right),
\qquad C>-\frac12 .
\label{eq:DeWitt-inner-long}
\end{align}
At a reference metric $\hat h(m)$ in a fixed conformal class, an
infinitesimal variation decomposes orthogonally in complex moduli
coordinates as
\begin{align}
\delta h_{ab}
&=2\,\delta\phi\,h_{ab}
+(Pv)_{ab}  +\sum_{i=1}^{n_g}
\left(\delta m^i\,\mu^{(i)}_{ab}
+\delta\bar m^{\bar i}\,\bar\mu^{(\bar i)}_{ab}\right),
\label{eq:metric-decomposition-long}
\end{align}
where
\begin{align}
(Pv)_{ab}
&=\nabla_av_b+\nabla_bv_a-h_{ab}\nabla_cv^c,
\nonumber\\
P^\dagger\mu^{(i)}&=0,\qquad h^{ab}\mu^{(i)}_{ab}=0.
\end{align}
Here $n_g=3g-3$ for $g>1$, $n_1=1$, and $n_0=0$.
The tensors $\mu^{(i)}$ span the transverse-traceless representatives
of the moduli tangent space. The Faddeev--Popov identity cancels the
integral along $v$ and the Weyl orbit against the gauge volume in
Eq.~\eqref{eq:polyakov-full-long}. The determinant of $P$ is
exponentiated by a traceless antighost $b^{ab}$ and a vector ghost
$c^a$,
\begin{equation}
\det{}'P
=\int\mc D b\,\mc D c\,
\exp\!\left[-\frac{1}{2\pi}
\int_\Sigma\dd\mu_h\,b^{ab}(Pc)_{ab}\right].
\label{eq:FP-ghost-long}
\end{equation}
The components of $b$ and $\bar b$ along the cokernel of $P$ are
saturated by
\begin{equation}
\prod_{i=1}^{n_g}
(b,\mu^{(i)})_h(\bar b,\bar\mu^{(\bar i)})_h ,
\label{eq:b-moduli-insertions-long}
\end{equation}
with the standard reduction by conformal Killing vectors at genus zero
and one. In complex coordinates Eq.~\eqref{eq:FP-ghost-long} is the
$(b,c)$ action of weights $(2,-1)$ and central charge $-26$.

Applying $Y=\mc I_hX$ at finite cutoff separates the area factor and the geometric Jacobian:
\begin{align}
\mc D_hX\,\e^{-S_s[X;h]}
&=[\Area[h]/\ell^2]^{-N/2}\mc J_s[h]\,
\mc D_hY\,\e^{-S_1[Y;h]},
\nonumber\\
\mc J_s[h]&=\mathfrak D[h]^{-N/2},
\qquad N=(s-1)D .
\label{eq:gauge-Jacobian-step-long}
\end{align}
On Eq.~\eqref{eq:fixed-area-slice-long} the area factor is unity, and the geometric factor transforms as
\begin{equation}
\frac{\mc J_s[e^{2\phi}\hat h]}{\mc J_s[\hat h]}
=\exp\!\left[\frac{N}{24\pi}
S_L[\phi;\hat h]\right].
\label{eq:J-Weyl-long}
\end{equation}
The combined $D$-boson and ghost measures contribute
\begin{equation}
\exp\!\left[\frac{D-26}{24\pi}
S_L[\phi;\hat h]\right].
\label{eq:Yghost-Weyl-long}
\end{equation}
The two Weyl weights add to
\begin{equation}
N+D-26=sD-26,
\label{eq:anomaly-sum-long}
\end{equation}
Collecting the moduli insertions, determinant density, local fields,
and Liouville factor gives Eq.~\eqref{eq:gaugefixed-long}, with
\begin{equation}
\dd m\equiv
\prod_{i=1}^{n_g}\dd^2m^i\,(b,\mu^{(i)})_{\hat h}
(\bar b,\bar\mu^{(i)})_{\hat h}.
\end{equation}
At $sD=26$ its Liouville coefficient vanishes.

\section{Flat-torus determinant}\label{app:torus}

For the unit-area metric in Sec.~\ref{sec:genusbrst}, the Fourier
modes $\psi_{m,n}=\exp[2\pi\ii(m\sigma^1+n\sigma^2)]$ give the
eigenvalues in Eq.~\eqref{eq:torus-eigenvalues-long}. First isolate
the line $m=0$. Zeta regularization and
$\zeta_{\mathrm R}(0)=-1/2$,
$\prod_{n=1}^\infty{}^\zeta n=\sqrt{2\pi}$ yield
\begin{align}
\prod_{n\ne0}^{\zeta}\lambda_{0,n}
&=\left(\frac{4\pi^2}{\tau_2}\right)^{2\zeta_{\mathrm R}(0)}
\prod_{n=1}^{\infty}{}^\zeta n^4
=\tau_2 .
\label{eq:torus-mzero-long}
\end{align}
For fixed $m\ne0$, the regularized number of $n$ modes is
$1+2\zeta_{\mathrm R}(0)=0$, so the prefactor
$4\pi^2/\tau_2$ drops out. The Weierstrass sine product gives
\begin{align}
\mc P_m
&\equiv\prod_{n\in\mathbb Z}^{\zeta}
\left[(n-m\tau_1)^2+(m\tau_2)^2\right]\nonumber\\
&=2\!\left[\cosh(2\pi|m|\tau_2)
-\cos(2\pi m\tau_1)\right]\nonumber\\
&=\e^{2\pi|m|\tau_2}|1-q^{|m|}|^2 ,
\qquad q=\e^{2\pi\ii\tau}.
\label{eq:torus-fixedm-long}
\end{align}
Multiplying the $m$ and $-m$ sectors and using
$\zeta_{\mathrm R}(-1)=-1/12$ gives
\begin{align}
\prod_{\substack{m\ne0\\ n\in\mathbb Z}}^\zeta\lambda_{m,n}
&=\prod_{m=1}^{\infty}{}^\zeta
\e^{4\pi m\tau_2}|1-q^m|^4\nonumber\\
&=\e^{4\pi\tau_2\zeta_{\mathrm R}(-1)}
\prod_{m=1}^{\infty}|1-q^m|^4\nonumber\\
&=\left|q^{1/24}\prod_{m=1}^{\infty}(1-q^m)\right|^4
=|\eta(\tau)|^4 .
\label{eq:torus-nonzerom-long}
\end{align}
Combining Eqs.~\eqref{eq:torus-mzero-long} and
\eqref{eq:torus-nonzerom-long},
\begin{equation}
\det{}'_\zeta A_\tau
=C\,\tau_2|\eta(\tau)|^4 ,
\label{eq:torus-det-app-long}
\end{equation}
where $C$ is independent of $\tau$. The Epstein zeta function
\begin{equation}
\zeta_{A_\tau}(z)
=(4\pi^2)^{-z}
\sum_{(m,n)\ne(0,0)}
\frac{\tau_2^z}{|m\tau-n|^{2z}}
\end{equation}
has value $-1$ at $z=0$ and derivative
\begin{equation}
\left.\frac{\dd}{\dd z}\right|_{z=0}
\sum_{(m,n)\ne(0,0)}
\frac{\tau_2^z}{|m\tau-n|^{2z}}
=-2\log\!\left(2\pi\sqrt{\tau_2}\,|\eta(\tau)|^2\right),
\end{equation}
and yields Eq.~\eqref{eq:torus-det-app-long}. If the torus has area $\mc A$, its eigenvalues are
$\mc A^{-1}\lambda_{m,n}$. Since $\zeta_A(0)=-1$ at genus one,
\begin{equation}
\det{}'A_{\mc A,\tau}
=C\,\mc A\,\tau_2|\eta(\tau)|^4,
\qquad
\frac{\det{}'A_{\mc A,\tau}}{\mc A}
=C\,\tau_2|\eta(\tau)|^4.
\end{equation}
The second line is the area-normalized determinant in Eq.~\eqref{eq:torusdet-long}.

\section{BRST charge and state-resolved sewing}\label{app:brst}

For the local completion, the matter stress tensor in Eq.~\eqref{eq:Tm-completion-long} has central charge $c_m=D+N$. The ghost OPE is
\begin{equation}
b(z)c(w)\sim\frac{1}{z-w},
\end{equation}
and the ghost stress tensor is
\begin{equation}
T_{\gh}(z)=-2:b\partial c:(z)-:(\partial b)c:(z),
\qquad c_{\gh}=-26.
\end{equation}
The matter and ghost stress tensors obey
\begin{align}
T_m(z)T_m(w)&\sim \frac{c_m/2}{(z-w)^4}+\frac{2T_m(w)}{(z-w)^2}+\frac{\partial T_m(w)}{z-w},\\
T_{\gh}(z)T_{\gh}(w)&\sim \frac{c_{\gh}/2}{(z-w)^4}+\frac{2T_{\gh}(w)}{(z-w)^2}+\frac{\partial T_{\gh}(w)}{z-w}.
\end{align}
The oscillator form of the holomorphic charge is
\begin{align}
Q_B
&=\sum_{n\in\mathbb Z}c_{-n}
\left(L_n^m-\delta_{n0}\right) -\frac12\sum_{m,n\in\mathbb Z}(m-n)
{:}c_{-m}c_{-n}b_{m+n}{:}.
\label{eq:BRST-oscillator-long}
\end{align}
where the intercept is one and
\begin{equation}
[L_m^m,L_n^m]
=(m-n)L_{m+n}^m
+\frac{c_m}{12}(m^3-m)\delta_{m+n,0}.
\label{eq:matter-Virasoro-long}
\end{equation}
Contracting $j_B(z)j_B(w)$ and separating total derivatives gives
\begin{equation}
\mathop{\rm Res}_{z=w}\big[j_B(z)j_B(w)\big]
=-\frac{c_m-26}{12}\,c\partial^3c(w)+\partial(\cdots).
\label{eq:BRST-residue-long}
\end{equation}
The displayed total derivative integrates to zero, so double contour
integration gives Eq.~\eqref{eq:Q2-long}. On
Eq.~\eqref{eq:critical-N-long}, $c_m=D+N=26$ and the obstruction
vanishes in the $Y+\chi$ operator algebra.

Let $|a;p,r\rangle$ denote an oscillator basis, where
$p\in\mathbb R^D$ and $r\in\mathbb R^N$ are the momenta of $Y$
and $\chi$. With BPZ metric
$G_{ab}(p,r)=\langle a;-p,-r|b;p,r\rangle$, the matter
identity is
\begin{equation}
\mathbf1_m
=\sum_{a,b}\int\frac{\dd^Dp\,\dd^Nr}{(2\pi)^{D+N}}\,
|a;p,r\rangle
\bigl(G^{-1}(p,r)\bigr)^{ab}
\langle b;-p,-r| .
\label{eq:matter-resolution-long}
\end{equation}
Insertion of Eq.~\eqref{eq:matter-resolution-long} on a plumbing
cylinder gives
\begin{align}
\mc A_{\rm sew}
&=\sum_{a,b}\int\frac{\dd^Dp\,\dd^Nr}{(2\pi)^{D+N}}
\langle\mc O_L|a;p,r\rangle
\bigl(G^{-1}\bigr)^{ab} 
\left\langle b;-p,-r\left|
q^{L_0-c_m/24}\bar q^{\bar L_0-c_m/24}
\right|\mc O_R\right\rangle .
\label{eq:sewing-app-long}
\end{align}
After adjoining the ghost resolution and plumbing insertions, BRST
quartets cancel. The angular integral over
$\arg q$ imposes level matching, and the residues of physical poles
factorize on BRST cohomology. Taking the trace produces the
continuous-momentum factor and oscillator product in
Eq.~\eqref{eq:torus-product-long}.

The distinguished massless state, including its ghost dressing, is
\begin{equation}
|\Psi_{\varepsilon,k}\rangle
=c_1\bar c_1\,\varepsilon_{\mu\nu}
\alpha_{-1}^\mu\bar\alpha_{-1}^\nu
|k\rangle_Y\otimes|0\rangle_\chi .
\end{equation}
The $L_0,\bar L_0$ terms of
$(Q_B+\bar Q_B)|\Psi_{\varepsilon,k}\rangle=0$ give $k^2=0$, while
the remaining level-one terms give
$k^\mu\varepsilon_{\mu\nu}
=k^\nu\varepsilon_{\mu\nu}=0$. A ghost-number-one gauge parameter
$|\Lambda\rangle$ generates the BRST-exact variation
\(
\delta|\Psi\rangle=(Q_B+\bar Q_B)|\Lambda\rangle
\), which induces
\begin{equation}
\delta\varepsilon_{\mu\nu}
=k_\mu\xi_\nu+\widetilde\xi_\mu k_\nu .
\end{equation}
This reproduces the constraints and gauge equivalence stated in
Sec.~\ref{sec:observables}.

\section{Background-field counterterms}\label{app:bg}

Use the covariant background split
\begin{equation}
Y^\mu(\sigma)
=\operatorname{Exp}_{\bar Y(\sigma)}
\!\left(\sqrt{\alpha'}\,\xi(\sigma)\right)^\mu .
\label{eq:background-split-long}
\end{equation}
The tangent fluctuation $\xi^\mu$ transforms as a target vector. Let
\begin{align}
D_a\xi^\mu
&=\partial_a\xi^\mu+\Gamma^\mu{}_{\nu\rho}
\partial_a\bar Y^\nu\xi^\rho,\nonumber\\
\mc D_a\xi^\mu
&=D_a\xi^\mu+\frac{\ii}{2}\epsilon_a{}^b
H^\mu{}_{\nu\rho}\partial_b\bar Y^\nu\xi^\rho .
\label{eq:background-bundle-connection-long}
\end{align}
The bundle connection $\mc D_a$ absorbs the first-derivative terms in
the covariant expansion of Eq.~\eqref{eq:sigma-action-long}. Integration by parts gives
\begin{align}
S_\sigma^{(2)}
&=\frac1{4\pi}\int\dd^2\sigma\sqrt h\,
\xi_\mu\mc O^\mu{}_\nu\xi^\nu,
\nonumber\\
\mc O&=-\left(h^{ab}\mc D_a\mc D_b+\mc E\right).
\label{eq:background-quadratic-long}
\end{align}
For Eq.~\eqref{eq:background-quadratic-long}, the first heat coefficient is
$a_1=\mc E+\tfrac16R^{(2)}\mathbf1$. Completing the square in the
$H D\xi\,\xi$ term and retaining the $\nabla H\,\xi^2$ term gives
\begin{widetext}
\begin{equation}
\left.\tr a_1\right|_{\partial\bar Y\partial\bar Y}
=-h^{ab}\!\left(R_{\mu\nu}
-\frac14H_{\mu\rho\sigma}H_\nu{}^{\rho\sigma}\right)
\partial_a\bar Y^\mu\partial_b\bar Y^\nu
+\frac{\ii}{2}\epsilon^{ab}\nabla^\rho H_{\rho\mu\nu}
\partial_a\bar Y^\mu\partial_b\bar Y^\nu .
\label{eq:a1-target-long}
\end{equation}
\end{widetext}
Equivalently, the symmetric and antisymmetric coefficients are the
parts of the torsionful Ricci tensor
\begin{align}
R^{(-)}_{\mu\nu}
&=R_{\mu\nu}
-\frac14H_{\mu\rho\sigma}H_\nu{}^{\rho\sigma}
-\frac12\nabla^\rho H_{\rho\mu\nu},\nonumber\\
R^{(-)}_{(\mu\nu)}
&=R_{\mu\nu}
-\frac14H_{\mu\rho\sigma}H_\nu{}^{\rho\sigma},
\qquad
R^{(-)}_{[\mu\nu]}
=-\frac12\nabla^\rho H_{\rho\mu\nu}.
\label{eq:torsion-Ricci-long}
\end{align}

In $2-\varepsilon$ dimensions, the proper-time representation gives
\begin{align}
\frac12\Tr\log\mc O\big|_{\rm pole}
&=-\frac12\int_0^\infty\frac{\dd t}{t}
\Tr\,\e^{-t\mc O}\bigg|_{\rm pole}
=-\frac{1}{4\pi\varepsilon}
\int\dd^2\sigma\sqrt h\,
\tr a_1 .
\label{eq:background-heat-pole-long}
\end{align}
The target-dependent trace in Eq.~\eqref{eq:a1-target-long} therefore
yields
\begin{widetext}
\begin{equation}
\Gamma_{\rm pole}^{(1)}
=\frac{1}{4\pi\varepsilon}\int\dd^2\sigma\sqrt h\,
\left[
h^{ab}\!\left(R_{\mu\nu}
-\frac14H_{\mu\rho\sigma}H_\nu{}^{\rho\sigma}\right)
+\ii\epsilon^{ab}\!\left(-\frac12\nabla^\rho
H_{\rho\mu\nu}\right)
\right]
\partial_a\bar Y^\mu\partial_b\bar Y^\nu .
\label{eq:GB-pole-long}
\end{equation}
\end{widetext}
The remaining scalar term is the Euler pole
$-D(24\pi\varepsilon)^{-1}\int\dd^2\sigma\sqrt h\,R^{(2)}$.
The counterterm is $S_{\rm ct}=-\Gamma_{\rm pole}^{(1)}$. Minimal
subtraction gives
\begin{align}
\beta^G_{\mu\nu}
&=\alpha'\left(R_{\mu\nu}
-\frac14H_{\mu\rho\sigma}H_\nu{}^{\rho\sigma}\right)
+O(\alpha'^2),\nonumber\\
\beta^B_{\mu\nu}
&=-\frac{\alpha'}2\nabla^\rho H_{\rho\mu\nu}
+O(\alpha'^2).
\label{eq:ordinary-GB-beta-long}
\end{align}

The Fradkin--Tseytlin coupling contributes the improvement associated
with the target-space vector $2\alpha'\nabla^\mu\Phi$. Equivalently,
the Weyl variation is combined with the local field redefinition
generated by this vector. Hence
\begin{align}
\bar\beta^G_{\mu\nu}
&=\beta^G_{\mu\nu}+2\alpha'\nabla_\mu\nabla_\nu\Phi,\nonumber\\
\bar\beta^B_{\mu\nu}
&=\beta^B_{\mu\nu}
+\alpha'\nabla^\rho\Phi\,H_{\rho\mu\nu},
\label{eq:dilaton-improvement-long}
\end{align}
which gives Eq.~\eqref{eq:GBbeta-long}. Define the scalar
Euler--Lagrange combination
\begin{equation}
\mc E_\Phi
=R+4\nabla^2\Phi-4(\nabla\Phi)^2-\frac1{12}H^2
-\frac{2(sD-26)}{3\alpha'} .
\label{eq:Ephi-long}
\end{equation}
The scalar heat coefficient, the curvature vertex from expanding
$\Phi(Y)$, and the Wess--Zumino consistency relation give
\begin{equation}
\bar\beta^\Phi
=\frac14G^{\mu\nu}\bar\beta^G_{\mu\nu}
-\frac{\alpha'}4\mc E_\Phi+O(\alpha'^2).
\label{eq:betaPhi-Ephi-long}
\end{equation}
Substituting Eq.~\eqref{eq:GBbeta-long} into
Eq.~\eqref{eq:betaPhi-Ephi-long} yields
\begin{align}
\bar\beta^\Phi
&=\frac{sD-26}{6}
+\alpha'\left[
(\nabla\Phi)^2-\frac12\nabla^2\Phi 
-\frac1{24}H^2\right]+O(\alpha'^2),
\end{align}
which is Eq.~\eqref{eq:phibeta-long}. The free $\chi$ determinant adds
$N/6$ to the scalar Weyl term, while the tensor counterterms arise from
the $Y$ sigma model. Including the $D$ coordinates and ghosts gives
$(D+N-26)/6=(sD-26)/6$.

Variation of Eq.~\eqref{eq:target-effective-action-long} gives
\begin{align}
0&=R_{\mu\nu}
-\frac14H_{\mu\rho\sigma}H_\nu{}^{\rho\sigma}
+2\nabla_\mu\nabla_\nu\Phi
-\frac12G_{\mu\nu}\mc E_\Phi,\nonumber\\
0&=\nabla^\rho H_{\rho\mu\nu}
-2\nabla^\rho\Phi\,H_{\rho\mu\nu},\nonumber\\
0&=\mc E_\Phi .
\label{eq:target-EOM-app-long}
\end{align}
Using the last line in the first gives
$\bar\beta^G_{\mu\nu}=0$, while the second is
$\bar\beta^B_{\mu\nu}=0$. For
$G_{\mu\nu}=g_{\mu\nu}^{\rm E}$, $H=0$, and
$\Phi=V_\mu Y^\mu$, $\mc E_\Phi=0$ reduces to
$sD+6\alpha'V^2=26$.


\begin{thebibliography}{99}
\bibitem{Pippard1953}
A.~B.~Pippard,
``An experimental and theoretical study of the relation between magnetic field and current in a superconductor,''
Proc. R. Soc. Lond. A \textbf{216}, 547 (1953), \doilink{10.1098/rspa.1953.0040}.

\bibitem{LaNaveRMP2019}
G.~La Nave, K.~Limtragool, and P.~W.~Phillips,
``Fractional electromagnetism in quantum matter and high-energy physics,''
Rev. Mod. Phys. \textbf{91}, 021003 (2019), \doilink{10.1103/RevModPhys.91.021003}.

\bibitem{GubserNonlocalSigma2019}
\begin{sloppypar}
S.~S.~Gubser, C.~B.~Jepsen, Z.~Ji,\allowbreak\ B.~Trundy, and A.~Yarom,
``Nonlocal nonlinear sigma models,''
JHEP \textbf{09}, 005 (2019),
\doilink{10.1007/JHEP09(2019)005}.
\end{sloppypar}

\bibitem{HeydemanNonlocalQED2020}
M.~Heydeman, C.~B.~Jepsen, Z.~Ji, and A.~Yarom,
``Renormalization and conformal invariance of non-local quantum electrodynamics,''
JHEP \textbf{08}, 007 (2020), \doilink{10.1007/JHEP08(2020)007}.

\bibitem{LaNavePhillips2019}
G.~La Nave and P.~Phillips,
``Anomalous dimensions for boundary conserved currents in holography via the Caffarelli--Silvestre mechanism for $p$-forms,''
Commun. Math. Phys. \textbf{366}, 119 (2019), \doilink{10.1007/s00220-019-03292-z}.

\bibitem{HartnollKarch2015}
S.~A.~Hartnoll and A.~Karch,
``Scaling theory of the cuprate strange metals,''
Phys. Rev. B \textbf{91}, 155126 (2015), \doilink{10.1103/PhysRevB.91.155126}.

\bibitem{GouterauxKiritsis2011}
B.~Gout\'eraux and E.~Kiritsis,
``Generalized holographic quantum criticality at finite density,''
JHEP \textbf{12}, 036 (2011), \doilink{10.1007/JHEP12(2011)036}.

\bibitem{Karch2014}
A.~Karch,
``Conductivities for hyperscaling violating geometries,''
JHEP \textbf{06}, 140 (2014), \doilink{10.1007/JHEP06(2014)140}.

\bibitem{DiazGiusti2018}
V.~A.~Diaz and A.~Giusti,
``Fractional Bosonic Strings,''
J. Math. Phys. \textbf{59}, 033509 (2018), \doilink{10.1063/1.5021776}.

\bibitem{Diaz2020}
V.~A.~Diaz,
``Virasoro Algebra and Asymptotic Symmetries from Fractional Bosonic Strings,''
\arxivlink{2009.11316}.

\bibitem{BasaClassification}
B.~Basa, G.~La Nave, and P.~W.~Phillips,
``Classification of nonlocal actions: Area versus volume entanglement entropy,''
Phys. Rev. D \textbf{101}, 106006 (2020),
\doilink{10.1103/PhysRevD.101.106006}, \arxivlink{1907.09494}.

\bibitem{Seeley1967}
R.~T.~Seeley,
``Complex powers of an elliptic operator,''
Proc. Symp. Pure Math. \textbf{10}, 288 (1967), \href{https://www.ams.org/books/pspum/010/}{AMS}.

\bibitem{ReedSimon}
M.~Reed and B.~Simon,
\textit{Methods of Modern Mathematical Physics. I: Functional Analysis}
(Academic Press, New York, 1980).

\bibitem{Balakrishnan1960}
A.~V.~Balakrishnan,
``Fractional powers of closed operators and the semigroups generated by them,''
Pacific J. Math. \textbf{10}, 419 (1960), \doilink{10.2140/pjm.1960.10.419}.

\bibitem{CS2007}
L.~Caffarelli and L.~Silvestre,
``An extension problem related to the fractional Laplacian,''
Commun. Partial Differ. Equations \textbf{32}, 1245 (2007), \doilink{10.1080/03605300600987306}.

\bibitem{FrassinoPanella2019}
A.~M.~Frassino and O.~Panella,
``Quantization of nonlocal fractional field theories via the extension problem,''
Phys. Rev. D \textbf{100}, 116008 (2019), \doilink{10.1103/PhysRevD.100.116008}.

\bibitem{Polyakov1981}
A.~M.~Polyakov,
``Quantum Geometry of Bosonic Strings,''
Phys. Lett. B \textbf{103}, 207 (1981), \doilink{10.1016/0370-2693(81)90743-7}.

\bibitem{Alvarez1983}
O.~Alvarez,
``Theory of Strings with Boundaries: Fluctuations, Topology, and Quantum Geometry,''
Nucl. Phys. B \textbf{216}, 125 (1983), \doilink{10.1016/0550-3213(83)90490-X}.

\bibitem{Vassilevich2003}
D.~V.~Vassilevich,
``Heat kernel expansion: user's manual,''
Phys. Rept. \textbf{388}, 279 (2003), \doilink{10.1016/j.physrep.2003.09.002}.

\bibitem{David1988}
F.~David,
``Conformal Field Theories Coupled to 2D Gravity in the Conformal Gauge,''
Mod. Phys. Lett. A \textbf{3}, 1651 (1988), \doilink{10.1142/S0217732388001975}.

\bibitem{DistlerKawai1989}
J.~Distler and H.~Kawai,
``Conformal Field Theory and 2D Quantum Gravity or Who's Afraid of Joseph Liouville?,''
Nucl. Phys. B \textbf{321}, 509 (1989), \doilink{10.1016/0550-3213(89)90354-4}.

\bibitem{Polchinski}
J.~Polchinski,
\textit{String Theory, Vol. 1}
(Cambridge University Press, Cambridge, 1998), \doilink{10.1017/CBO9780511816079}.

\bibitem{GuillarmouRhodesVargas2019}
C.~Guillarmou, R.~Rhodes, and V.~Vargas,
``Polyakov's formulation of $2d$ bosonic string theory,''
Publ. Math. IHES \textbf{130}, 111 (2019), \doilink{10.1007/s10240-019-00109-6}.

\bibitem{Quillen1985}
D.~Quillen,
``Determinants of Cauchy--Riemann operators over a Riemann surface,''
Funct. Anal. Appl. \textbf{19}, 31 (1985), \doilink{10.1007/BF01086022}.

\bibitem{BPZ1984}
A.~A.~Belavin, A.~M.~Polyakov, and A.~B.~Zamolodchikov,
``Infinite conformal symmetry in two-dimensional quantum field theory,''
Nucl. Phys. B \textbf{241}, 333 (1984), \doilink{10.1016/0550-3213(84)90052-X}.

\bibitem{KatoOgawa1983}
M.~Kato and K.~Ogawa,
``Covariant Quantization of String Based on BRS Invariance,''
Nucl. Phys. B \textbf{212}, 443 (1983), \doilink{10.1016/0550-3213(83)90680-6}.

\bibitem{KobaNielsen1969}
Z.~Koba and H.~B.~Nielsen,
``Manifestly crossing-invariant parametrization of $n$-meson amplitude,''
Nucl. Phys. B \textbf{12}, 517 (1969),
\doilink{10.1016/0550-3213(69)90071-6}.

\bibitem{Friedan1980}
D.~Friedan,
``Nonlinear Models in $2+\epsilon$ Dimensions,''
Phys. Rev. Lett. \textbf{45}, 1057 (1980), \doilink{10.1103/PhysRevLett.45.1057};
Annals Phys. \textbf{163}, 318 (1985), \doilink{10.1016/0003-4916(85)90384-7}.

\bibitem{Callan1985}
C.~G.~Callan, D.~Friedan, E.~J.~Martinec, and M.~J.~Perry,
``Strings in Background Fields,''
Nucl. Phys. B \textbf{262}, 593 (1985), \doilink{10.1016/0550-3213(85)90506-1}.

\bibitem{FradkinTseytlin1985}
E.~S.~Fradkin and A.~A.~Tseytlin,
``Effective Field Theory from Quantized Strings,''
Phys. Lett. B \textbf{158}, 316 (1985), \doilink{10.1016/0370-2693(85)91190-6}.
\end{thebibliography}
\end{document}